\documentclass[usenatbib]{mn2e}
\usepackage{graphicx}                                            
\usepackage{times}

\title[Echoes from the companion]{Echoes from the companion star in Sco X-1}

\author[T. Mu\~noz-Darias et al.]
{T. Mu\~noz-Darias$^1$ \thanks{E-mail: tmd@iac.es},
I.G. Mart\'\i{}nez-Pais$^{1,2}$, J. Casares$^{1}$, V.S. Dhillon$^{3}$
\newauthor
 T.R. Marsh$^{4}$, R. Cornelisse$^{1}$, D. Steeghs$^{4,5}$, P.A. Charles$^{6,7}$\\
$^1$ Instituto de Astrof\'{\i}sica de Canarias, 38200 La Laguna, Tenerife, Spain\\
$^2$ Departamento de Astrof�ica, Univ. de La Laguna, E-38206 La Laguna, Tenerife, Spain\\
$^3$ Dept. of Physics \& Astronomy, Univ. of Sheffield, Sheffield S3 7RH, UK\\
$^4$ Dept. of Physics, Univ. of Warwick, Coventry CV4 7AL, UK\\
$^5$ Harvard-Smithsonian Center for Astrophysics, Cambridge, MA 02138, USA\\
$^6$ South Africa Astronomical Observatory, P.O. Box 9. Observatory 7935, South Africa\\
$^7$ University of Southampton, Southampton, SOB17 1BJ, UK}
\begin{document}

\maketitle

\begin{abstract}
We present simultaneous X-ray (RXTE) and optical (ULTRACAM) narrow band (Bowen blend/HeII and nearby continuum) observations of Sco X-1 at 2-10 Hz time resolution. We find that the Bowen/HeII emission lags the X-ray light-curves with a light travel time of $\sim 11-16$s which is consistent with reprocessing in the companion star. The echo from the donor is detected  at orbital phase $\sim 0.5$ when Sco X-1 is at the top of the Flaring Branch. 
Evidence of echoes is also seen at the bottom of the Flaring Branch but with time-lags of 5-10 s which 
are consistent with reprocessing in an accretion disc with a radial temperature profile. We discuss the implication of our results for the orbital parameters of Sco X-1. 
%Our time-delays suggest that either the inclination angle is higher than $50^{\circ}$ or the mass of the primary is heavier than $1.4 M_{\odot}$.
\end{abstract}


\section{Introduction} \label{introduction}
\noindent Low mass X-ray binaries (LMXBs) are interacting binaries containing a low mass donor star
transferring matter onto a neutron star (NS) or black hole.
Mass accretion takes place through an accretion disc, and with temperatures approaching $\sim 10^8$ K,
such systems are strong X-ray sources. The mass transfer rate supplied by the donor star, $\dot{M}_2$, is driven 
by the binary/donor evolution and for $\dot{M}_2>\dot{M}_{crit}\sim 10^{-9} M_{\odot} yr^{-1}$ results in persistently 
bright X-ray sources. In these binaries optical emission is dominated by reprocessing of the powerful X-ray luminosity in the gas around the compact object which usually swamp  the spectroscopic features of the weak companions stars. Only in exceptional cases, such as long-period
LMXBs with much brighter evolved donors (e. g. Cyg X-2) is a complete orbital solution possible
(\cite{casa98}; hereafter CCK98). In this scenario, dynamical studies have classically been
restricted to the analysis of X-ray transients during quiescence, where the intrinsic
luminosity of the donor dominates the light spectrum of the binary (e.g. \cite{char04}).
%The accretion disc subtends the largest solid angle as viewed by the X-ray source and, therefore, is responsible for most of theirradiated light.
\subsection{Fluorescence Emission from Donor Stars}
Fortunately, this situation has recently changed thanks to the discovery
of the narrow emission components arising from the donor star in the prototypical persistent
LMXB Sco X-1 \cite{stee02}. High resolution spectroscopy revealed many narrow high-excitation
emission lines, the most prominent associated with HeII $\lambda4686$ and NIII $\lambda\lambda$4634-41 /
CIII $\lambda\lambda$4647-50 at the core of the broad Bowen blend. In particular, the NIII lines are powered by fluorescence resonance through
cascade recombination which initially requires EUV seed photons of HeII
Ly${\alpha}$.  These very narrow (FWHM $\leq 50$km s$^{-1}$) components move in antiphase with
respect to the wings of HeII $\lambda$4686, which
approximately trace the motion of the compact star. Both properties
(narrowness and phase offset) imply that these components originate in
the irradiated face of the donor star. This work represented the first
detection of the companion star in Sco X-1 and opened a new window for
extracting dynamical information and thereby deriving mass functions in a
population of $\sim$ 20 LMXBs with established optical counterparts.
We now know that this property is not peculiar to Sco X-1 but is a feature
of persistent LMXBs, as demonstrated by the following examples:\\
(i)~Doppler tomography of NIII $\lambda4640$ in the eclipsing ADC ({\it accretion disc corona}) pulsar 2A 1822-371 reveals a 
compact spot at the position of the companion star and with velocity $K_{em}=300 \pm 8$ km s$^{-1}$ (\citealt{casa03}). 
Moreover, the application of the K-correction
to this $K_{em}$ velocity, combined with other system parameters, strongly points to the presence 
of a neutron star  with mass $1.6-2.3M_{\odot}$ in this system~(\citealt{mcm05}).\\
(ii)~Radial velocities of narrow Bowen lines in the black hole candidate
GX339-4, detected during the 2002 outburst, led to a mass function in
excess of $5.8 M_{\odot}$, and hence provided the first dynamical proof 
that the compact object is indeed a black hole(\citealt{hynes03}).\\
(iii)~Sharp NIII $\lambda$4640 Bowen emission has been detected in 4U 1636-536,
4U 1735-444 (\citealt{c06}), Aql X-1 (\citealt{remon07}) and the transient millisecond pulsar XTE J1814-338(\cite{casa04}), which lead to
donor velocity semi-amplitudes in the range 200-300 km s$^{-1}$.

\subsection{Echo-Tomography}
Echo-tomography is an indirect imaging technique which uses time delays
between X-ray and UV/optical light-curves as a function of orbital phase
in order to map the reprocessing sites in a binary (\cite{obrien02}; hereafter OB02).
The optical light curve can be simulated by the
convolution of the (source) X-ray light curve with a transfer function which
encodes information about the geometry and visibility of the reprocessing
regions. The transfer function~(TF) quantifies the binary response to the
irradiated flux as a function of the lag time and it is expected to have two main components:
the accretion disc and the donor star. The latter is strongly dependent on
the inclination angle, binary separation and mass ratio and, therefore,
can be used to set tight constraints on these fundamental parameters.
Successful echo-tomography experiments have been performed on several X-ray
active LMXBs using X-ray and broad-band UV/optical light-curves. The results
indicate that the reprocessed flux is dominated by the large
contribution of the accretion disc (e.g. \citealt{hynes98};hereafter H98, OB02, \citealt{obrien04} and
\citealt{hynes05}) which dilutes the reprocessed signal arising from the companion. Only \cite{midd81} and \cite{dav75} have reported the detection of reprocessed pulsations in the X-ray binaries 4U 1626-67 and Her X-1 respectively. They only used optical data but the Doppler phase modulation of the coherent pulses leads to reprocessing of the X-ray pulses on the companion star. More recently, \cite{hynes06} 
have reported some evidence for a reprocessed signal from the companion 
through the detection of orbital phase dependent echoes in EXO 0748-676.\\
Exploiting emission-line reprocessing rather than broad-band
photometry has two potential benefits: a) it amplifies the response of
the donor's contribution by suppressing most of the background
continuum light (which is associated with the disc); b) since the lines are formed in an optically thin medium, the 
reprocessing time is almost instantaneous and hence the response is
sharper (i.e. only smeared by geometry) and the transfer function
easier to compute (see \citealt{Munoz05}; hereafter MD05). Taking advantage of this, we decided to
undertake an echo-tomography campaign on the brightest LMXB of all,
Sco X-1, in order to search for reprocessed signatures of the donor using simultaneous X-ray and Bowen/HeII emission line light-curves.\\ 
\subsection{Background to Sco X-1 / V818 Sco}
Sco X-1 is the prototype LMXB and also the brightest persistent X-ray source in the sky and has been the
target of detailed studies since its discovery (\citealt{giacconi62}). An orbital period of 18.9 hr was reported
by \citealt{gotti75} from photographic photometry of the V$\sim$13 optical counterpart, V818 Sco, using plates spanning over 85 years. 
The optical B-band light curve shows a 0.13 mag amplitude
modulation (\citealt{Augusteijn92}), which is interpreted as due to the changing visibility
of the inner-hemisphere of the X-ray heated companion. Radial velocity curves of HeII$\lambda$4686
and HI confirmed the orbital period and indicated that the inferior
conjunction of the emission line regions are close to the photometric
minimum and, hence, they must originate near the compact object (e.g. \citealt{LaSala85}). \citealt{Bradshaw99} measured the trigonometric 
parallax of Sco X-1 using VLBA radio observations and deduced a distance of
$2.8\pm0.3$ kpc. Further observations revealed the presence of twin radio
lobes, which are inclined at an angle of $44^{\circ} \pm 6^{\circ}$ relative
to the line of sight (\citealt{Fomalont01}; hereafter F01). Furthermore, Sco X-1 is classed 
as a $Z-source$ due to the Z-shape patterns that the spectral variations trace in the X-ray hard/soft colour-colour diagram. 
The source also exhibits kHz QPOs (\citealt{VdK96}) but no X-ray bursts have been observed so far.\\
Possible correlated optical and X-ray variability has been studied by using broad band observations in both spectral ranges. 
In particular, \cite{Ilovaisky80} and \cite{Petro81} present evidence for correlated variability during flaring episodes. They measured time-lags 
consistent with reprocessing in the accretion disc. Here we present the first detection of delayed echoes arising from 
the companion star in Sco X-1 together with other echoes more consistent with reprocessing in the accretion disc. 
The echoes are found during different X-ray states within the Z-shape pattern. 
We discuss the implications of our results for reprocessing theory and the system parameters of Sco X-1.
\begin{table}
\begin{center}
\caption{Sco X-1 optical observing Log}
\label{ucamlog}
\begin{tabular}{c c c c c}
\hline
Date(2004) & $T_{EXP}$(s) & \textit{Seeing}(") & $\phi_{orbital}$ & S/N\\
\hline
17 May & 0.1 & 0.5-1 & 0.07-0.35 & 60\\
18 May & 0.25-0.5 & 1-3 & 0.34-0.73 & 50-100\\
19 May & 0.3 & 1-2& 0.73-0.95 & 90\\
\hline
\end{tabular}
\end{center}
\end{table}
\section{Observations and Data Reduction} \label{Data}
Simultaneous X-ray and optical data of Sco X-1 were obtained on the nights
of 17-19 May 2004.  The full 18.9 hr orbital period was covered
in 12 snapshots, yielding 16.1 ks of X-ray data simultaneous with optical
photometry in 3 different bands.

\subsection{X-ray data}
The X-ray data were obtained with the Proportional Counter Array (PCA) onboard the Rossi X-ray Timing Explorer (RXTE) satellite (\citealt{Jahoda2006}). 
Only 2 PCA detectors ($2$ and $5$) were used and the pointing offset was
set to 0.71$^{\circ}$ due to the brightness of Sco X-1. The data were analysed
using the FTOOLS software and the times corrected to the solar barycentre.
The STANDARD-1 mode, with a time resolution of 0.125s, was used for the variability analysis.
Additionally, the PCA energy bands  $b$($2$-$4$ keV), $c$($4$-$9$ keV) and $d$($9$-$20$ keV) of the 
STANDARD-2 mode data, with a time resolution of 16s, were used for the spectral analysis.
%The minimum count rate  of $5000$ counts s$^{-1}$ implies that background subtraction was not necessary. 
Sco X-1 was observed during 15 RXTE windows of 16-31 minutes length spread over the three nights and yielding 20 ks of data. 
About 80\% of these observations were also covered with simultaneous optical data.  
\begin{figure}
\includegraphics[width=8cm]{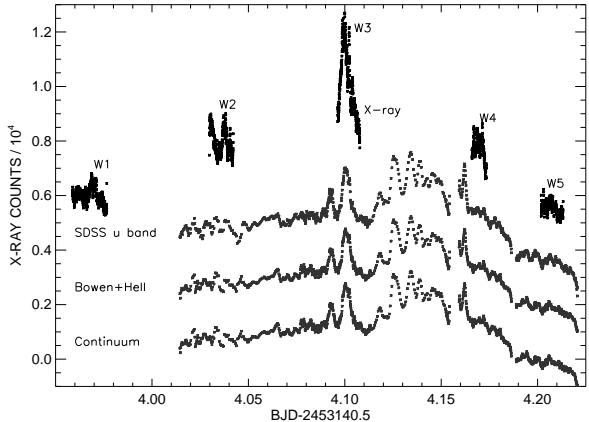}
\caption{Simultaneous RXTE and ULTRACAM (SDSS $u'$, Bowen+HeII and Continuum) observations of Sco X-1/V818 Sco obtained on the night of 18 May 2004.}
\label{data}
\end{figure}

\subsection{Optical data}
The optical data were obtained with ULTRACAM at the Cassegrain focus of the 4.2m William Herschel Telescope (WHT) at La Palma. ULTRACAM is a triple-beam CCD camera which uses two dichroics to split the
light into 3 spectral ranges: Blue ($<\lambda$3900),
Green($\lambda\lambda$3900-5400) and Red ($>\lambda$5400).
It uses frame transfer 1024x1024 E2V CCDs which are
continuously read out, and are capable of time resolution down to 2 milliseconds by
reading only small selected windows (see \citealt{dhi07} for details).
ULTRACAM is equipped with a standard set of $ugriz$ SDSS filters. However,
since we want to amplify the reprocessed signal from the companion, we decided to use two narrow
(FWHM =100 \AA) interference filters in the Green and Red channels,
centered at $\lambda_{\rm eff}$=4660\AA~ and $\lambda_{\rm eff}$=6000\AA. These filters
block out most of the continuum light and allow us to integrate two
selected spectral regions: the Bowen/HeII blend and a
featureless continuum, from which continuum-subtracted light-curves of the
high excitation lines can be derived. A standard SDSS $u'(3543$ \AA) filter was also mounted in the blue channel.
The images were reduced using the ULTRACAM pipeline software with bias subtraction and flatfielding. Star counts were extracted by adjusting the aperture radii to the seeing, and light-curves were obtained relative to a comparison
star which is 96 arcsecs NW of Sco X-1. Light-curves of the Bowen/HeII lines (hereafter B+HeII CS)
were computed by subtracting the Red (Continuum) from the Green(hereafter B+HeII) channels (see section 3.1.2).
The seeing was 1-1.5 arcsecs most of the time, except for the first two RXTE windows of 18 May when it degraded to over 3 arcsecs. Optical observations during the first 3 RXTE visits on 19 May were ruined by clouds.
The exposure time was initially set to 0.1s but was increased to 0.25s depending on weather conditions. Integrations of 0.5s had to be used for the second window on 18 May, when the seeing was worse. An observing log is presented in
Table \ref{ucamlog}.

\section{Light-curves}
In order to look for correlated variability we performed a comprehensive cross-correlation analysis by using the entire data set in blocks from 2 min to 30 min (i.e. one RXTE window). Significant cross-correlation peaks were found only on the night of 18 May, when the X-ray flux rose up to $1.3 \times 10^4$ counts s$^{-1}$, which is a factor $\sim 2$ higher than the mean flux displayed by the source during the whole campaign. During this night the system also showed long episodes of flaring activity (see fig \ref{data}). Unfortunately, during the first two RXTE visits the weather conditions were very poor and no useful optical data could be obtained. In this work we are going to focus on both the 3rd and 5th RXTE visits of May 18th where we detect clear episodes of correlated variability between the X-ray and optical data. We also note that some correlated variability appears in the 4th RXTE visit but in this case the correlation peaks are noisier and many uncorrelated features are also present in the data.

\subsection{Echo detection of the companion at superior conjunction}
\begin{figure}
  \includegraphics[width=8cm]{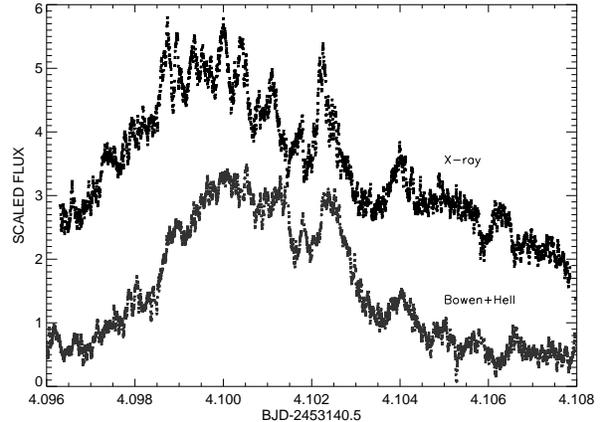}
  \caption{Zoom of W3 showing a 1000s stream of X-ray (above) and B+HeII non continuum subtracted (below) data. Both low and high frequency variability appear clearly correlated.}
  \label{w3}
\end{figure}

In fig. \ref{w3} we present a zoom of the large flare seen during the 3rd RXTE visit of 18 May (W3) together with the simultaneous B+HeII light curve. During W3, where a big flare occurs, the X-ray flux reached $1.3 \times 10^4$ counts s$^{-1}$, the maximum count rate in the whole campaign. We note significant X-ray variability with clear correlated counterparts in the optical data. According to the orbital ephemeris reported by SC02, W3 is centered at orbital phase 0.52 i.e. very close to the superior conjunction of the donor star, when the irradiated face presents its largest visibility with respect to us. Assuming a light path from the X-ray source to the center of the companion, the expected time delay ($\tau$) for reprocessing in the donor is:
\begin{equation}
\label{ret}
\tau=\frac{a}{c}(1-\sin{i}\cos{2 \pi \phi_{orb}})
\end{equation}
\begin{figure}
  \includegraphics[width=8cm]{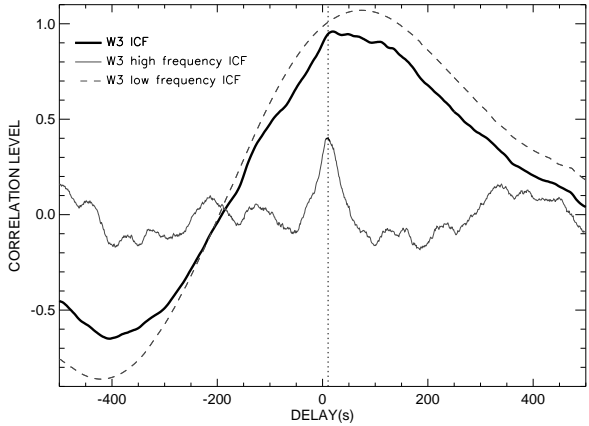}
  \caption{ICF analysis for W3. The thick solid line presents the ICF obtained between the X-ray and B+HeII emission. We also plot the ICF of the low (dashed line) and high frequency (thin solid line) variability after filtering the light curve with a step function at 0.002Hz. A clear peak corresponding to a delay of 10-11s (dotted line) is obtained in the latter case.}
  \label{c_w3}
\end{figure}
where $i$ is the inclination angle, $a$ the orbital separation, $\phi_{orb}$ the orbital phase and $c$ the velocity of light. Therefore, $\phi_{orb}\sim0.5$ is the phase when reprocessing in the companion star is expected to be at maximum. For instance, if we use  standard values for $a$($\sim 10$ lt-sec) and $i$($\sim 44^{\circ}$), we obtain $\tau \sim 17$s, which should be considered as an upper limit to $\tau$ since the reprocessing take place on the heated hemisphere of the donor and not in its center of mass.\\
\subsubsection{Cross-correlation analysis}
\begin{figure}
  \includegraphics[width=8cm]{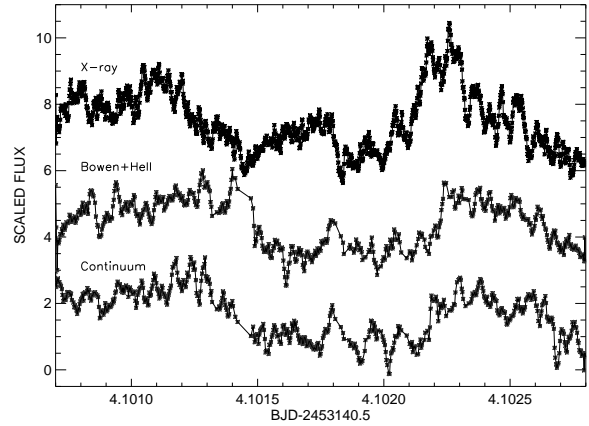}
  \caption{The 3 min interval, selected from W3, for the ICF analysis.}
  \label{b67}
\end{figure}

\begin{figure}
  \includegraphics[width=8cm]{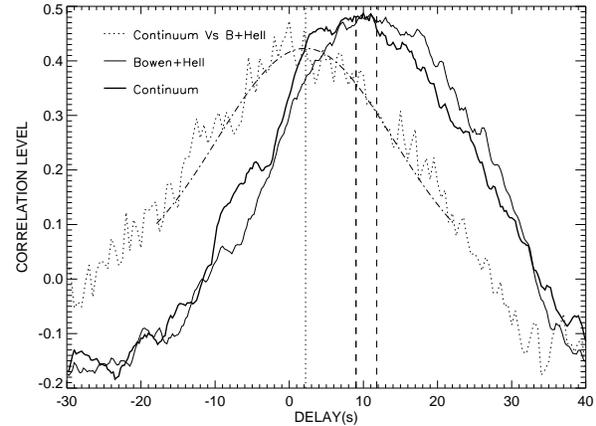}
  \caption{Continuum versus Bowen+HeII ICFs for the selected 3min block of W3. The solid lines show the correlations between the X-ray and both the B+HeII and Continuum emission. As a test we also plot (in dotted line) the ICF between the Continuum and Bowen+HeII data. A gaussian fit to the peak is over-plotted as the dotted-dashed line. The vertical lines mark the measured time-lag for each case: $11.8\pm0.1$s, $9.0\pm0.7$s and $2.2\pm0.3$s respectively}
  \label{cor_bvc}
\end{figure}

As a first step we have computed cross correlation functions between the X-ray and the B+HeII light-curves by using the entire data set obtained during W3. We have used a modified version of the ICF (Interpolation Correlation Function) method which is explained in great detail by \citealt{g&p87}. We also tested the DCF~(Discrete Correlation Function) method (\citealt{ek88}) but found no differences in their results. Fig. \ref{c_w3} shows the ICF obtained for W3 as the thick solid line, where a clear peak centered at 8-30 s is obtained. Since the light-curves in W3 show evidence for both high and low frequency variability we decided to filter, prior to computing the ICFs, in order to deconvolve the two components.To do this, we convolved the data with a $sinc$ function to isolate the low frequency component. The light-curve corresponding to the high frequency variability was obtained by subtracting the low frequency component light-curve from the original data.  The ICF computed for frequencies lower than 0.002 Hz, which roughly traces the profile of the big flare, is plotted in fig. \ref{c_w3} as the dashed line. This shows a strong and wide peak centered at \textbf{$\sim 40-80$s}. This time-lag is clearly longer than the reprocessing time scale~(RTS) which, for Sco X-1, is expected to be in the range $0-20$s. Hence, it cannot be associated with the light travel time alone and its origin will be  discussed in section 5. The high frequency($\geq 0.002$ Hz) analysis, on the other hand, yields a flat ICF (fig. \ref{c_w3} thin solid line) with a narrow peak centered at a lower delay. We have fitted this peak to both parabolic and gaussian functions finding a time-lag between~$10-11$s. This delay is consistent with reprocessing on the companion star (see section 5). Based on this result we decided to focus our analysis on short ($\sim 2-5$ min) data intervals dominated by high frequency variability. We obtain highly significant correlation peaks by choosing different data blocks during W3 especially during the second part of this window where strong correlated variability is present(see fig \ref{w3}). In many cases we find ICF time-lags of about 11s between the X-ray and B+HeII light-curves. For detailed analysis we finally selected a 3 min block where the correlation peaks are particularly significant. This interval is presented in fig. \ref{b67}. The resulting ICF function is shown in figure \ref{cor_bvc} as the thin line. A Gaussian fit to the peak yields a time lag of $11.8\pm0.1$s, where the error represents the formal error of the fit. In order to establish the confidence levels we have performed a Monte Carlo simulation where synthetic ICF were computed from a large population ($10^5$) of white noise time series (random, zero-mean normal distributions of data) with the same sampling as the optical data. The correlation of each random function with respect to the original X-ray data was computed and then a statistic was calculated. Thus a 3$\sigma$ level corresponds to the upper envelope which encompasses 99.7\% of the computed ICFs. For the selected interval we find a~$3\sigma$ confidence at a correlation level of 0.125 which is much lower than our ICF peak value.\\
The same analysis was also performed with the light-curves obtained using the SDSS~$u'$ and Continuum filters. The corresponding ICFs also show clear peaks centered at positive delays with correlation levels very close to 0.5. However, in these two cases the correlation peaks are wider and non-symmetric, and hence the error obtained from the Gaussian fits are larger. In figure \ref{cor_bvc} we also show the ICF for the Continuum data which lags the X-ray emission by $9.0\pm0.7$s. In the same figure we have over-plotted as the dotted line the ICF obtained between the Continuum and the B+HeII emission where we measure a delay of $2.2\pm0.3$s. This time-lag is consistent with the above results. On the other hand, we find that the $u'$ band emission lags the X-ray data by $9.5\pm0.7$s, which is also $\sim 1-3$s shorter than the delay obtained with the B+HeII filter.
\begin{figure}
  \includegraphics[width=8cm]{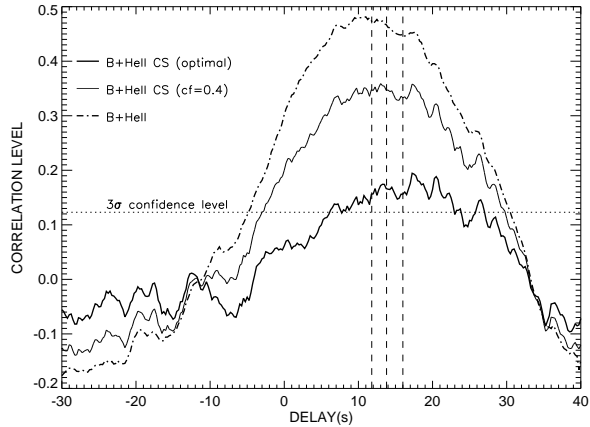}
  \caption{ICF for the selected 3min block of W3. The ICFs have been computed using the X-ray data and a B+HeII light curve obtained by subtracting from the green channel the Continuum data scaled by a factor $cf=0$ (i. e. no subtraction), 0.4 and 0.8 (optimal subtraction). The dotted line shows the $3\sigma$ confidence level and the dashed vertical lines time-lags of $11.8\pm0.1$s, $13.3\pm0.1$s and  $16.0\pm0.3$s respectively.}
  \label{cor_b67}
\end{figure}

\subsubsection{ICF analysis for continuum subtracted B+HeII light-curves}
SC02 pointed out that at least 10\% of the flux contained in the Bowen/HeII emission lines arise from the companion star. Thus, in an attempt to amplify the signal from the donor we decided to subtract the Red (Continuum) from the Green(B+HeII) channels. However, we have seen above that the Continuum is strongly correlated with the X-ray emission, and hence, the subtraction process will reduce the correlation level of the signal and will add noise. Furthermore, the continuum level will likely be different at the central wavelength of the red continuum filter and the Bowen spectral region. The ratio of continuum levels probably depends on the X-ray state of the source as well as perhaps the orbital phase since the companion contribution changes around the orbit. Therefore, we performed a test to determine the optimal amount of continuum to be subtracted in order to amplify efficiently the reprocessed signal from the companion. Table \ref{cfac} presents the ICF parameters obtained after subtracting a fraction $cf$ (with $cf$ in the range 0-1) of the red continuum from the B+HeII light curve. The table lists both the delay and the correlation level (obtained through a Gaussian fit) as a function of $cf$. It is clear that the higher is $cf$ (i. e. more continuum is subtracted) the lower the correlation level of the peak. On the other hand, the ICF peak smoothly shifts to longer delays for larger $cf$ values, as expected if more disc contribution is subtracted. We find that for $cf$ values greater than $\sim 0.8-0.85$ the ICF functions become noisier and secondary peaks start to appear with correlation levels comparable to the main peak. Since for $cf = 0.80$ the correlation level of the peak is still significant with the 99.99\% confidence level (i. e.~$4\sigma$) we decided to take this value as our best estimate of the continuum subtraction level for the ICF analysis. We call this 'optimally' subtracted lightcurve 'B+HeII CS', although it should be noted that this definition is somewhat arbitrary and could depend on our signal-to-noise. We also note that the $1\sigma$ errors quoted in the table are formal and they are unrealistically small. Therefore we decided to choose a more conservative delay in the range 14.3-16.3s, corresponding to $cf$=0.6-0.8, when the $1\sigma$ errors start to rise.\\

 In order to best estimate the actual amount of subtracted continuum it is necessary to take into account both the different throughputs of the filters and the quantum efficiency of the detector, together with the spectral distribution of Sco X-1. For the case of  a flat $f_\lambda$ distribution between $\sim 4600$ \AA~and $6000$ \AA, we estimate that $\sim 95$\% of the continuum level has been subtracted. On the other hand, if we assume the spectral distribution reported by \cite{scha89} the actual subtraction level for $cf=0.8$ would be $\sim60-70\%$. However, it is clear that the spectral distribution must depend on the X-ray state and the orbital phase, and therefore simultaneous optical spectroscopy would be needed to constrain the continuum level at the Bowen blend spectral region at the time of the observations. Note that only the narrow components of the Bowen Blend arise from the companion, and hence the disc contribution can not be totally removed by subtracting continuum light from the B+HeII emission. Since the delay obtained for the B+HeII CS is $\sim 3-4$s higher than for the non subtracted data, this analysis strongly suggests that the companion signal dominates over the disc emission when most of the continuum contribution is removed from the B+HeII data. This is illustrated in fig. \ref{cor_b67} where we present the ICFs for $cf$=0, 0.4 and 0.8. The plot clearly shows the monotonic shift of the peaks to longer delays when $cf$ increases. Note that we obtain similar results by fitting parabolic functions to the ICF peaks instead of Gaussians.
\begin{table}
\begin{center}
\caption{Time lag and correlation level versus Continuum subtraction factor}
\label{cfac}
\begin{tabular}{c c c}
\hline
Continuum factor ($cf$) & Correlation level & Mean delay$\pm 1\sigma$(s)\\
\hline
0.00 & 0.48 & $11.8\pm0.1$\\
%0.05 & 0.47 & $11.9\pm0.1$\\
0.10 & 0.45 & $12.1\pm0.1$\\
%0.15 & 0.44 & $12.2\pm0.1$\\
0.20 & 0.42 & $12.4\pm0.1$\\
%0.25 & 0.41 & $12.6\pm0.1$\\
0.30 & 0.39 & $12.8\pm0.1$\\
%0.35 & 0.37 & $13.1\pm0.1$\\
0.40 & 0.35 & $13.3\pm0.1$\\
%0.45 & 0.33 & $13.5\pm0.1$\\
0.50 & 0.31 & $13.8\pm0.1$\\
%0.55 & 0.29 & $14.1\pm0.1$\\
{\bf0.60} & {\bf0.26} & ${\bf14.5\pm0.2}$\\
%0.65 & 0.24 & $15.0\pm0.2$\\
{\bf0.70} & {\bf0.22} & ${\bf15.4\pm0.2}$\\
%0.75 & 0.19 & $15.8\pm0.2$\\
{\bf0.80} & {\bf0.18} & ${\bf16.0\pm0.3}$\\
0.85 & 0.15 & $16.5\pm0.4$\\
\hline
\end{tabular}
\end{center}
\end{table}
\begin{figure}
  \includegraphics[width=8cm]{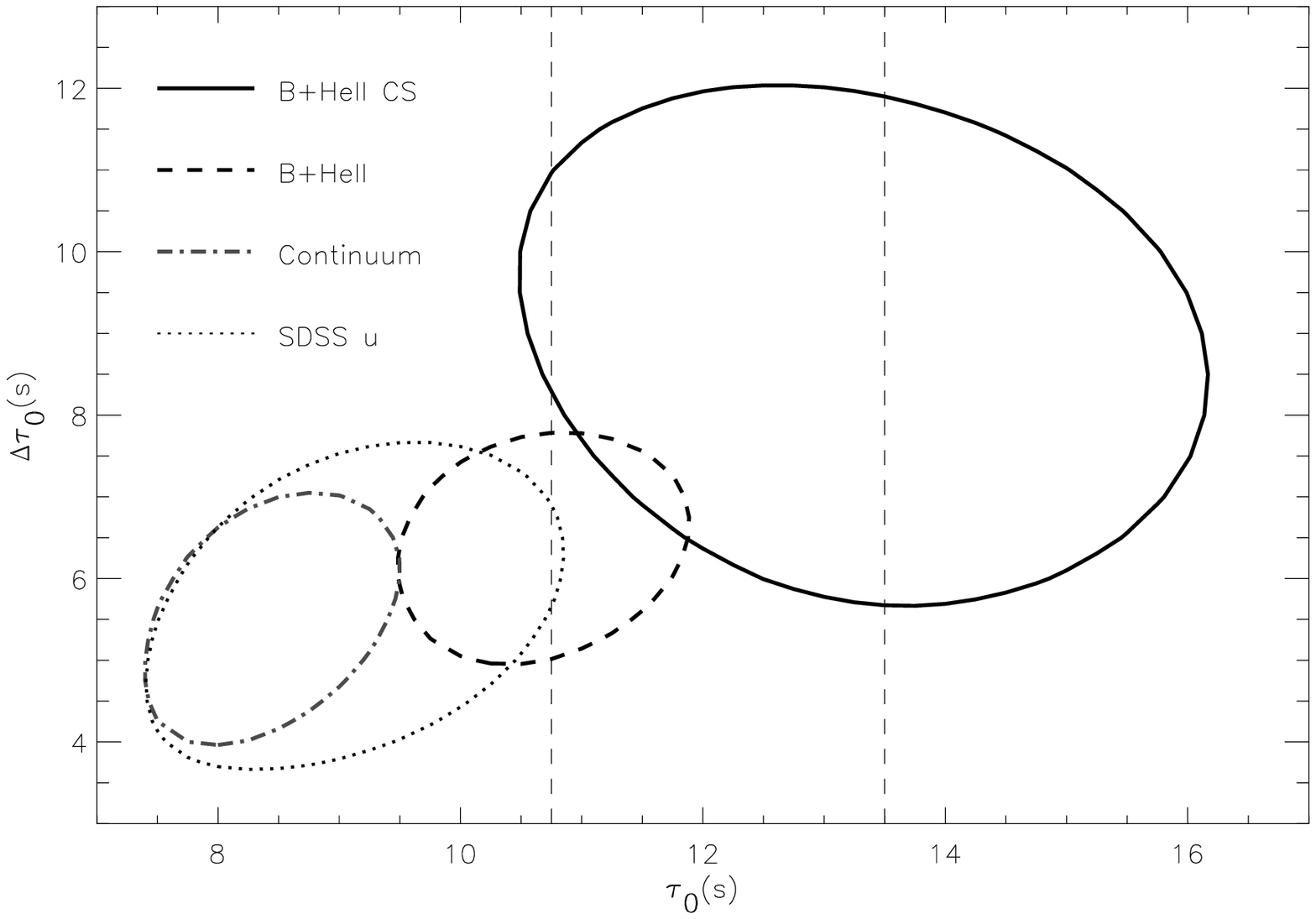}
  \caption{Gaussian Transfer Function fit for the SDSS $u'$, Continuum, B+HeII and B+HeII CS data. We show the $1\sigma$ and $2.6\sigma$ contours for all the cases. Both TF parameters, $\tau_0$ and $\Delta\tau_0$, are highly constrained by the fit.}
  \label{gauss_b67}
\end{figure}
\begin{figure}
  \includegraphics[width=8cm,height=10cm]{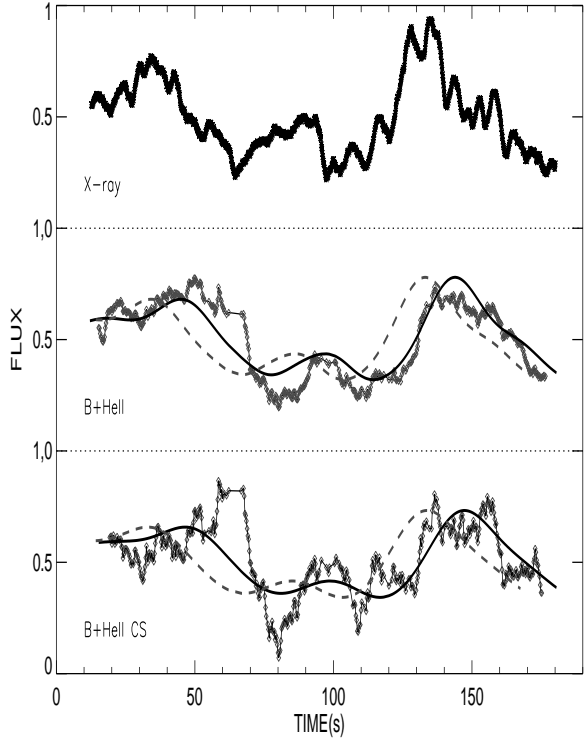}
  \caption{X-ray light-curve (upper panel) and its convolution with $TF(\chi_{min}^2)$ superimposed on the B+HeII (middle panel) and B+HeII CS (lower panel) data. In both cases the convolution between the X-ray data and $TF(\chi_{min}^2)$ for $\tau_0=0$ is over-plotted as a dashed line. The light-curves have been smoothed for clarity.}
  \label{cur_g_b67}
\end{figure}     
\subsubsection{Fitting transfer functions}
In a second step we decided to fit the ULTRACAM light-curves with the results of convolving the RXTE data with a set of synthetic transfer functions. For simplicity we have used Gaussian TFs with two free parameters which represent the mean delay ($\tau_0$) and the standard deviation of the delay ($\Delta\tau$):
\begin{equation}
\label{tf}
TF(\tau)=A e^{\frac{1}{2}(\frac{\tau-\tau_0}{\Delta\tau})^2}
\end{equation}
where A is a normalization constant. This TF, which was first used in H98, is different to that computed by MD05 for the specific case of reprocessing in emission lines. However, we think that this is more appropriate since  the data quality is not good enough to justify the use of such an accurate model which considers separately the companion and disc contributions. We have applied this technique to all data sets (i.e. B+HeII, B+HeII CS, SDSS $u'$ band and Continuum) and find $\tau_0$ values completely consistent with those obtained in the cross-correlation analysis. Fig. \ref{gauss_b67} shows the minima of the $\chi_{\nu}^2$ distribution for all the filters together with the $1\sigma$ confidence level. A summary of the results is presented in Table \ref{g_res}.\\
 \begin{table}
\begin{center}
\caption{TF and ICF  analysis for W3}
\label{g_res}
\begin{tabular}{c | c c c}
\hline
 &  ICF$\pm1\sigma$(s) & $\tau_0\pm1\sigma$(s) & $\Delta\tau_0\pm1\sigma$(s)\\
\hline
B+HeII CS & 14.3-16.3  & $13.5\pm3.0$&$8.5\pm3.5$ \\
B+HeII &$11.8\pm0.1$  & $10.75\pm1.25$& $6.25\pm1.5$\\
SDSS $u'$ &$9.5\pm0.7$ &$9.0\pm1.5$&$5.5\pm2.0$ \\
Continuum&$9.0\pm0.7$ &$8.5\pm1.0$ &$5.5\pm1.5$ \\
\hline
\end{tabular}
\end{center}
\end{table}

This new independent analysis confirms the time-lags obtained with the ICF method for the three channels. We have also applied this technique to the continuum subtracted data as we did for the ICF study. Again, we find that the mean delays shift to higher values when the continuum is subtracted. However, for $cf>0.5-0.6$ the $1\sigma$ contours become considerably large as well. Although the maximum delay is obtained for $cf=0.8$ we have selected $cf=0.7$ as our preferred subtraction level for the TF analysis since the lower limit of the  $1\sigma$ contour hits its maximum value and hence is more constraining. This results in $\tau_0=13.5\pm3.0$. On the other hand, the $\Delta\tau_0$ values are in the range 5-7 seconds for the three optical bands, whereas for the subtracted light-curve we get $\Delta\tau_0=8.5\pm3.5$s. Since the TF fitting is based on a Gaussian smoothing of the X-ray light-curve, the method tends to remove the uncorrelated features by increasing $\Delta\tau_0$ and then does not attempt to fit the high frequency variability. However the fits successfully reproduce the main behaviour of the optical light-curves. We show in Fig. \ref{cur_g_b67} the fits for the optical (B+HeII and B+HeII CS) data for  $TF(\chi_{min}^2)$. We also show (dashed lines) that $\tau_0=0$ is clearly inconsistent with the optical observations.

\subsection{Correlated variability during W5}
We also find evidence for correlated variability during the last RXTE window on the night of 18 May (W5). W5 is centered at $\phi_{orb}= 0.66$ and during the observation the X-ray flux remained at $6000$ counts s$^{-1}$ (i. e. half of the count rate in W3). In figure \ref{b8} we show a 15 min block of data during W5 where the correlated variability is clearly detected. Following Sect. 3.1, we have analysed the data using both ICF and TF. 
\subsubsection{ICF analysis}
\begin{figure}
  \includegraphics[width=8cm]{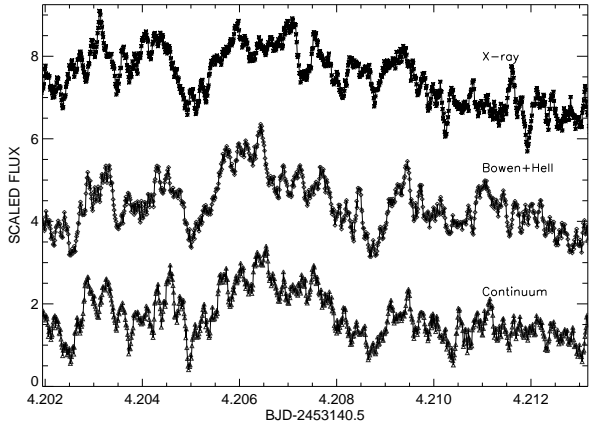}
  \caption{1000s of X-ray (above), B+HeII (middle) and Continuum (below) data during W5.}
  \label{b8}
\end{figure}

\begin{figure}
  \includegraphics[width=8cm]{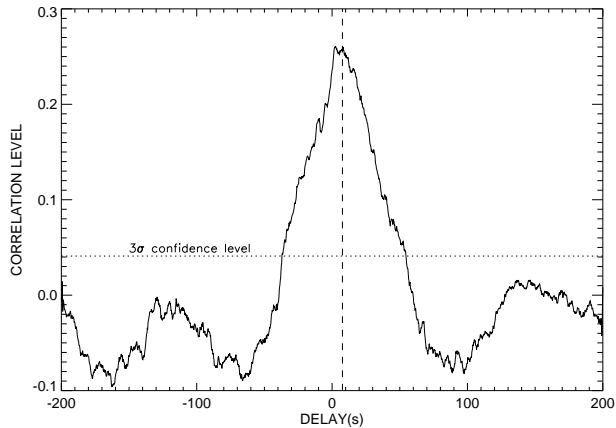}
  \caption{ICF obtained during W5 for the X-ray and B+HeII data. The dotted line shows the $3\sigma$ confidence level and the dashed one the resulting time-lag of $7.6\pm0.4$s.}
  \label{c_b8}
\end{figure}

The light curve obtained during W5 (see fig. \ref{b8}) is essentially flat and only high frequency ($\geq 0.002$ Hz) variability is present. For the ICF analysis we have considered several data intervals within W5 and finally selected a $\sim 6$~min block which corresponds to the first part of the data presented in figure \ref{b8} (from day no. 4.202 to 4.206) and the correlation is best seen. Note that the same delays are obtained by using the entire W5 data set, although for this 6 min segment the correlation peaks are more symmetric and hence the delays are better extracted from a gaussian fit. The ICF obtained for the three ULTRACAM bands (i.e. SDSS $u'$, B+HeII and Continuum) clearly show highly significant peaks centered at $5.5\pm0.2$s, $7.6.\pm0.4$s and $9.4\pm0.4$s respectively as obtained through Gaussian fits.  The ICF function for the B+HeII case is shown in figure \ref{c_b8}. The shapes of the ICFs are very similar for the three bands and have a peak correlation level of $\sim 0.25$. This value is much higher than the $3\sigma$ confidence level obtained for this data interval, which is at 0.04.\\
In order to obtain light-curves of the emission lines we have subtracted the red (Continuum) from the green(B+HeII) channels in the same way as we did for W3.  We find that the peak's correlation level also decreases when $cf$ increases but no shift to higher delays is observed in this case. In contrast to the results for W3, the ICF  peaks tend to slightly shorter delays when the continuum is subtracted. This is consistent with the time-lags presented above for W5.
\subsubsection{TF analysis}
\begin{figure}
  \includegraphics[width=8cm]{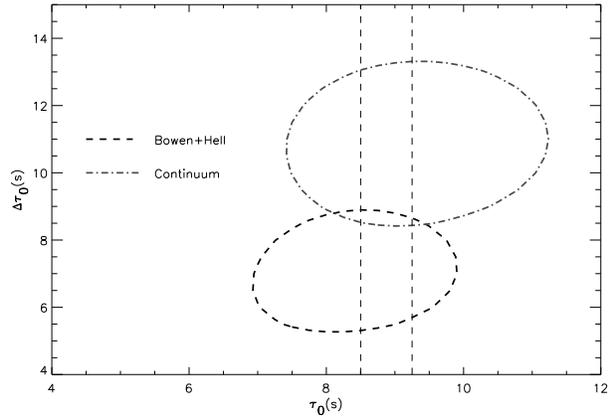}
  \caption{Gaussian Transfer Function fit for both B+HeII and Continuum data from W5. The dashed lines mark the $\tau_0$ found for each case. The contours give the $1\sigma$ confidence level for $\tau_0$ and $\Delta\tau_0$.}
  \label{g_b8}
\end{figure}

In the same way as we did for W3 we have computed synthetic optical light-curves by convolving the X-ray data with a Gaussian TF (see eq. \ref{tf}). Figure \ref{g_b8} presents the $\chi^2$ map obtained from the entire W5 window(see fig. \ref{b8}). We obtained $\tau_0= 8.5\pm1.5$s and $\tau_0= 9.25\pm1.75$s for the B+HeII and Continuum data respectively.  These results are consistent with those obtained with the ICF method. In table \ref{rb8} we summarize the results obtained with both techniques.\\
 \begin{table}
\begin{center}
\caption{TF and ICF analysis for W5}
\label{rb8}
\begin{tabular}{c | c c c}
\hline
 & ICF$\pm1\sigma$(s) & $\tau_0\pm1\sigma$(s) & $\Delta\tau_0\pm1\sigma$(s)\\
\hline
B+HeII &$7.6\pm0.4$ & $8.5\pm1.5$ & $7.0\pm2.0$\\
SDSS $u'$ &$5.5\pm0.2$ &$7.5\pm3.5$&$18.5\pm4.0$ \\
Continuum&$9.4\pm0.4$ &$9.25\pm1.75$ &$10.5\pm2.5$ \\
\hline
\end{tabular}
\end{center}
\end{table}

\begin{figure}
  \includegraphics[width=8cm]{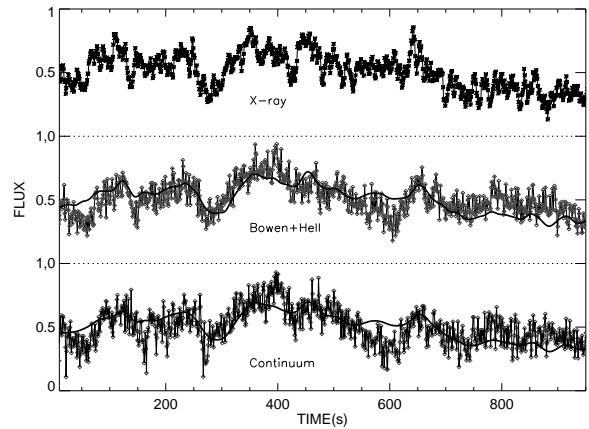}
  \caption{X-ray emission (top panel) convolved with $TF(\chi_{min}^2)$ and plotted on top of B+HeII (middle panel) and Continuum(lower panel) data. The light-curves have been smoothed for clarity.}
  \label{c_g_b8}
\end{figure}

Although the fit is poor we obtain $\tau_0= 7.25\pm3.5$s for the $u'$ band which is consistent with the ICF analysis. On the other hand, we get  $\Delta\tau_0=18.5\pm4.0$s and $\Delta\tau_0=10.5\pm2.5$ for the $u'$ and Continuum bands respectively. These values are larger than those obtained for W3 and this is  probably  due to the presence of features in the X-ray data which are not present in the optical light-curves (see section 5.). We also obtain $\Delta\tau_0=7.0\pm7.0$s for B+HeII and this is consistent with the results for W3 within the errors.
Figure \ref{c_g_b8} shows the fits of the convolved X-ray light-curves overplotted on the optical data. Although there are small features which are not correlated, the fit successfully reproduces the main behaviour of the optical data.

\begin{figure}
  \includegraphics[width=8cm,height=12cm]{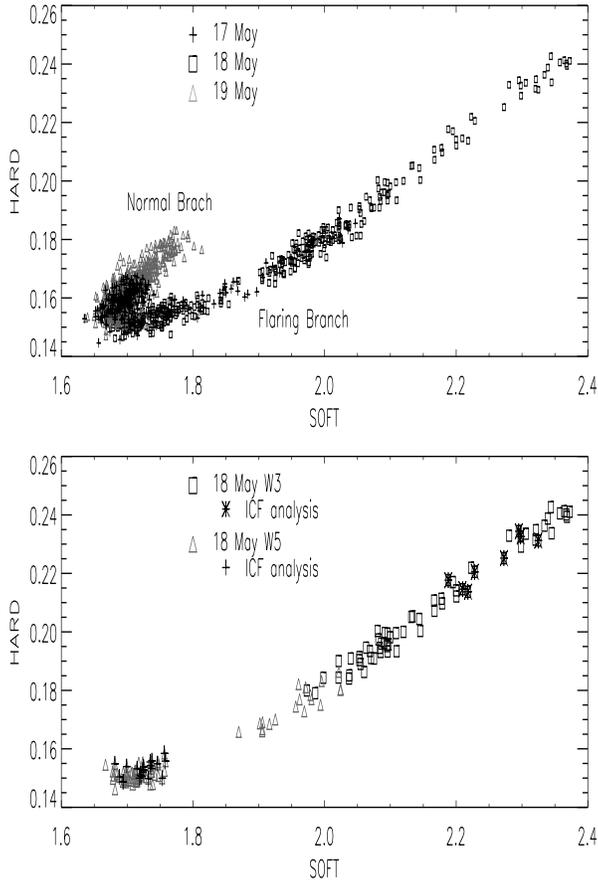}
  \caption{Upper panel: colour-colour diagram obtained for the whole observing campaign. During the first and third night the source was in the NB whereas it moved to the FB during the second night, where correlated variability is found. In the lower panel we separate the data of W3 and W5 finding that Sco X-1 was in a different X-ray state in the two epochs where we detect the echoes.}
\label{coco}
\end{figure}
\section{Colour-Colour Diagram analysis}
The broad band X-ray spectral variations of Z-sources are usually studied by building colour-colour diagrams (\citealt{HVdK89}). We have used the PCA energy bands $b$($2$-$4$ keV), $c$($4$-$9$ keV) and $d$($9$-$20$ keV) in order to  compute the Hard($d/c$) and Soft($c/b$) X-ray colours. The colour-colour diagram obtained from the data  of the whole campaign is presented in the upper panel of figure \ref{coco}. Sco X-1 was at the bottom of the Normal Branch (NB) during most of the 17 May data but moved towards the Flaring Branch (FB) in the last RXTE visit, showing a 50 percent increase in flux. On 18 May Sco X-1 was in the FB and exhibited large amplitude variability, with large flares similar to those seen during W3.  On 19 May the system stayed on the NB again. As we showed in the light curve analysis, we have only detected correlated variability during the second night when the system moved across the FB. In the lower panel of fig. \ref{coco} we present the colour-colour diagram for the night of 18 May and separate the W3 and W5 data using different symbols. We have also marked the intervals within these windows where the ICF analysis has been performed. It is clear that the ICF peak obtained during W3 is associated with the hardest X-ray emission, whereas during W5 (where lower delays are detected for the B+HeII emission) the system was close to the transition between the FB and NB. This study suggests that reprocessing associated with the companion star is most easily detected when the X-ray flux is at maximum and the spectra are harder. Note that although the spectra is harder in W3, the flux corresponding to both the soft and hard X-ray emission components is at maximum during this window.  

\section{discussion}
We have found clear evidence for optical echoes to the X-ray emission when Sco X-1 is moving across the FB. 
We have detected two different delay time scales by using both the low and high frequency variations present in the both X-ray and optical data. The low frequency analysis results in large delays of $\sim 40-70$s which are clearly not consistent with the RTS expected for Sco X-1 if we only assume light-travel times. Such large delays have been previously seen by other authors. In particular a $\sim 30$s delay was reported by I80 in Sco X-1. \citealt{Cominsky87} studied the local diffusion time-scale during  X-ray reprocessing. They found that 50\% of the X-ray photons are reemitted within the first $\sim 0.2$s, although a long tail of delays up to 10s is found as well. Since the light travel times within Sco X-1 should be lower than 15-18s, it is clear that the time-lags obtained for the low frequency variability are not compatible with X-ray reprocessing on any possible site within the binary. Although the nature of these long time-lags is unclear we note that they are marginally consistent with the time scale for re-adjustment to thermal equilibrium in the accretion disc, which is expected to be about a few minutes (e. g. \citealt{ap}).\\

\begin{figure}
  \includegraphics[width=8cm]{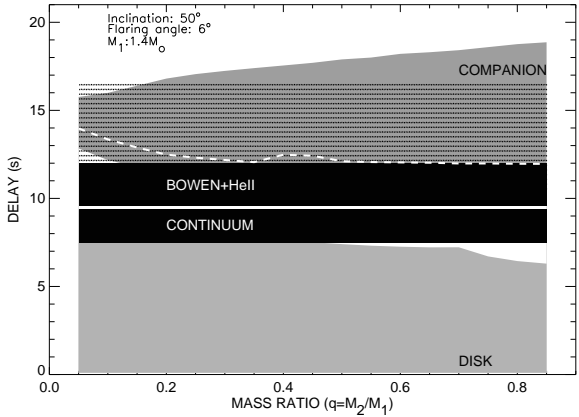}
  \caption{Expected range of delays for both the companion and the accretion disc considering $i=50^{\circ}$, $M_1=1.4M_{\odot}$ and the orbital parameters of Sco X-1. The black region shows the $1\sigma$ delays obtained for the B+HeII, and Continuum data. The shadowed region represents the range of delays obtained by subtracting the Continuum from the B+HeII light-curve. The white dashed line is the highest probability delay for the companion according to MD05 and using $\alpha=6^{\circ}$.}
\label{dvq}
\end{figure}

On the other hand, the higher frequency variability component yields shorter delays ($10-11$s) which are indeed consistent with X-ray reprocessing within the binary. In particular, during  W3 we have found that the SDSS u band, the B+HeII emission and the Continuum data lag the X-ray emission with light travel times of $\sim 9$s, $\sim 11$s and $\sim 9$s respectively. In order to measure these delays we have used both ICF and TF methods finding excellent agreement between the two techniques. 
The case of the B+HeII emission is especially interesting since SC02 demonstrated that it is partially associated with emission arising from the companion star in Sco X-1. This emission is produced by fluorescence resonance through cascade recombination of HeII Ly${\alpha}$ photons in a low density medium, and hence we expect almost instantaneous reprocessing times. Fig. \ref{dvq} shows the range of expected time-delays for both the companion and the accretion disc as a function of the mass ratio ($q=M_2/M_1$). In this simulation we have considered the TF presented in MD05 for the standard orbital parameters~($i=50^{\circ}$, $M_1=1.4M_{\odot}$) of Sco X-1 and $\phi_{orb}=0.52$. 
The time-lags of $\sim 10-11$s obtained for the B+HeII (unsubtracted) data during W3 lies at the lower edge of the companion region, and shows that part of this emission must arise from reprocessing on the donor star. Hence, this time-delay represents the first direct detection of delayed echoes from the companion star.\\ 
 
In an attempt to isolate the emission from the B+HeII lines we subtracted the Continuum data from the green channel and we have applied both the ICF and TF techniques to the obtained light-curve. 
%After performing the analysis we have selected $cf=0.8$ as the subtraction level for the ICF method whereas for the TF technique $cf=0.7$ gives the most restrictive time-delay measurement. According to this latter method
By using the TF technique we obtain a time-lag of $13.5\pm3.0$s which is perfectly accommodated in the companion region (fig. \ref{dvq}). On the other hand, the ICF method yields a time-lag between $14.3-16.3$s (see table \ref{cfac}). According to the TFs computed in MD05, this time-lag is larger than expected for the maximum response of the companion for each value of $q$ (white dashed line). Here, we have assumed a disc flaring angle of $\alpha=6^{\circ}$ but the discrepancy is still present if we use larger $\alpha$ values. This suggests that some of the values used for the orbital parameters of Sco X-1 might not be correct. Furthermore, if we consider that the disc extends up to the truncation radius computed in \citealt{Packy} we find that the time-delay of $\sim 9$s obtained for the Continuum is also slightly larger than expected for the outer disc. But this difference could be reconciled if the Continuum light were a combination of both reprocessed emission from the companion and the accretion disc. This is supported by the orbital modulation seen in the optical light-curve, which is associated with heating of the companion (e. g.  SC02). However this is in contradiction with the $\sim 9$s lag also measured for this filter during W5. Since W5 is centered at $\phi_{orb}=0.66$ the delay associated with reprocessing on the companion must be lower than for W3~($\phi_{orb}=0.52$), and hence the measured time-lag should also be lower. We propose two possible scenarios in order to better accommodate our measured ICF time-delays with the model computed for Fig. \ref{dvq}:\\
(i) The inclination angle of the system is $\sim 60^{\circ}$, significantly higher than  the $i=44 \pm 6^{\circ}$ reported by F01. This latter result was obtained by assuming that the jets are perpendicular to the orbital plane of the binary. This assumption is difficult to test but there is evidence for misalignments in some X-ray binaries (see \citealt{Fender} and references therein).\\
(ii) The neutron star (NS) has a mass of $\sim 1.8 M_{\odot}$. Note that there are evidents of heavier~($M_{NS}>1.6M_{\odot}$) neutron stars in some LMXBs (e. g. Cyg X-2 ;CCK98, V395 Car; \citealt{Tariq04}, \citealt{Jonker05} or X1822-371;MCM05)

%To date, the most accurate neutron star mass determinations have been obtained by studying radio pulsars. These studies indicate that the mass of the neutron star at formation must be close to $\sim 1.35 \pm 0.04 M_{\odot}$ (\citealt{vK01}). However, recent results have revealed evidence for heavier~($M_{NS}>1.6M_{\odot}$) neutron stars in Cyg X-2 (CCK98), the old white dwarf + pulsar J0751+1807~(\citealt{Nice05}),  V395 Car (\citealt{Tariq04}; \citealt{Jonker05}) , X1822-371 (MCM05) and Aql X-1 (\citealt{remon06}).\\
Obviously, a combination  of the two scenarios can also solve the discrepancy i.e. $i\sim 55^{\circ}$ and NS mass $\sim 1.6 M_{\odot}$. We stress that this interpretation relies on the result derived through the continuum subtraction analysis for the ICF method. Alternatively, we obtain a more conservative result by using the TF method which results in time lags consistent with the standard parameters of Sco X-1.\\

The $\sim 9$s delay obtained for the SDSS $u'$ band during W3 is much larger than measured during W5~($\sim6$s). This is similar to the behaviour of the B+HeII emission where a $\sim 11$s lag was measured during W3 and $\sim 8$s during W5, but different to the Continuum filter, where a $\sim 9$s delay is observed both in W3 and W5. SC02 have detected narrow components within the Bowen blend which arise from the companion star and also have reported the presence of many other high ionization lines at $\lambda \leq 4700$ in the spectrum of Sco X-1. They showed that at least one of these emission lines also arises from the companion star. According to the UV spectra of Sco X-1 presented in S89 there are many high excitation lines within the spectral range covered by the SDSS $u'$ band. In particular, this range covers three Bowen OIII lines at $\lambda\lambda 3030-3750$, which probably also arise from the companion. The contribution of these lines to the $u'$ band light could explain the larger time-delays obtained for this band during W3.\\
 \citealt{Willis80} pointed out that the intensity of the high excitation emission lines present in the UV spectrum ($\lambda\lambda 1100-1800$) of Sco X-1 increases by a factor 1.5-3.2 when the source moves towards a high UV state. These authors associated this high UV state with a high energy X-ray state by using simultaneous UV/X-ray data (see also \citealt{White76}). Based on this study we propose that while Sco X-1 stays at the bottom of the FB (W5) the intensity of the high excitation lines is not large enough to dominate over the continuum signal which arises from the accretion disc. And we note that the distribution of time-lags is correlated with wavelength for W5. This is consistent with a scenario where optical emission arises through reprocessing in an accretion disc with a radial temperature profile. Moreover, when we subtract the Continuum from the B+HeII data in W5 the delays do not shift to higher values. This shows that the B+HeII emission is dominated by the same signal as the Continuum during W5.\\
On the other hand, W3 is centered at orbital phase $\sim 0.5$ when the inner hemisphere of the companion presents its largest visibility with respect to us. Our colour-colour diagram shows that at this time the source was at the top of the FB, which resulted in  higher X-ray emission (factor $\sim 2$ with respect to W5) and a harder X-ray spectrum. According to W80, during this high energy state the intensity of the Bowen emission lines could have increased enough to dominate over the continuum signal and yielding the echo detection obtained for the W3 B+HeII data. On the other hand, we note that during the first part of W3 the system was also at the top of the FB but no time-lag consistent with the companion was detected. This shows that there are still some gaps in our knowledge of when and how it is possible to detect the reprocessing signal from the companion.\\
For the case of the continuum subtracted data, where the errors are larger, we find that the accuracy provided by the ICF method is much higher than  obtained through the TF technique. Nevertheless, the latter has proved very useful to determine time-lags and their corresponding uncertainties for the three bands considered. \citealt{Mc03} applied this technique to the sets of Sco X-1 data used by I80 and P81. They measured $\tau_0=8.0\pm0.8$s and $\Delta\tau_0=8.6\pm1.3$s for the Johnson $B$ band data. This filter covers our B+HeII narrow band and these results are consistent with ours during W5. The TF method also provides information about the time-delay distribution, which is encoded in the parameter $\Delta\tau_0$. During W3 we have measured $\Delta\tau_0 \sim 5-6$s in the three bands whereas it increases after subtracting the Continuum  from the B+HeII emission. These results are consistent with the time-lag distribution expected for reprocessing within the binary. Although the increase in $\Delta\tau_0$ detected for the B+HeII CS data is consistent with the higher time-lag measured for this light-curve, we note that  $\Delta\tau_0$ is also sensitive to the presence of uncorrelated features in the X-ray data. Since the TF fitting is based on a Gaussian smoothing of the X-ray light-curve, the method tends to remove the non-correlated features by increasing $\Delta\tau_0$. This clearly happens in the TF analysis of the SDSS $u'$ band during W5 where uncorrelated features are present.  

\section{Conclusion}
We, for the first time, have isolated emission 
line contributions from the mass donor star that respond to X-ray variations 
with a delay that is consistent with light-travel time bewteen the X-ray 
source near the neutron star and the irradiated face of the donor star that 
directly contributes to the emission. The correlated signal from the companion is found at an orbital phase close to the superior conjunction of the donor, when a large flare occurs and the system stays at the top of the FB. In particular, the B+HeII emission is dominated by the reprocessed emission from the companion which increases the time-lag by 25\% in comparison to that obtained in a featureless continuum at $\lambda_{eff} = 6000$. After subtracting the continuum emission from the B+HeII light-curve we amplify the signal from the donor and measure time lags in the range 10.5-16.5s by using both the TF and ICF techniques. We also note that for the latter case we have detected $4\sigma$ significant ICF peaks centered at 14-16s which suggests that either the orbital inclination or the NS mass could be higher than the standard values assumed. We stress here that a second detection of a B+HeII echo from the companion at a different orbital phase would directly constrain the inclination angle of the system. A more detailed study using HeII Ly${\alpha}$ and Bowen line light-curves could probably increase the signal from the companion, thereby allowing  a more accurate determination of the system parameters in Sco X-1.\\

 %or EXO 0748-676 (Ozel et al. 2006).
\section*{Acknowledgments}
ULTRACAM is supported by PPARC grant PP/D002370/1. JC acknowledges support from the Spanish Ministry of Science and
Technology through the project AYA2002-03570. TRM was supported by a PPARC Senior Research Fellowship 
during the course of this work. DS acknowledges a Smithsonian Astrophysical Observatory Clay
Fellowship as well as support through NASA GO grant NNG04G014G.

 %"In order to check the significance of the correlations, we performed a Montecarlo test by generating a number of white noise time series (random,zero-mean normal distributions of data) with the same sampling than theoriginal moving series. The correlation of each random function with respect to the original fixed function was computed and then a statistic was performed among the obtained correlation functions. Thus, for example, a 4$\sigma$ level corresponds to a value for which 99.99\% of the computed correlations have values below it. We can thus say that the probability of having a peak exceeding this level in the correlation of a random time series with the first function is 0.01\%."

\bsp

\label{lastpage}

\end{document}